\documentstyle{cupconf}
\input{epsf.sty} 

\ifoldfss
\else
  \ifnfssone
    \newmathalphabet{\mathit}
      \addtoversion{normal}{\mathit}{cmr}{m}{it}
      \addtoversion{bold}{\mathit}{cmr}{bx}{it}
    \newmathalphabet{\mathcal}
      \addtoversion{normal}{\mathcal}{cmsy}{m}{n}
    \else
    \ifnfsstwo
    \fi
  \fi
\fi

\newbox\grsign \setbox\grsign=\hbox{$>$} \newdimen\grdimen \grdimen=\ht\grsign
\newbox\simlessbox \newbox\simgreatbox \newbox\simpropbox
\setbox\simgreatbox=\hbox{\raise.5ex\hbox{$>$}\llap
     {\lower.5ex\hbox{$\sim$}}}\ht1=\grdimen\dp1=0pt
\setbox\simlessbox=\hbox{\raise.5ex\hbox{$<$}\llap
     {\lower.5ex\hbox{$\sim$}}}\ht2=\grdimen\dp2=0pt
\setbox\simpropbox=\hbox{\raise.5ex\hbox{$\propto$}\llap
     {\lower.5ex\hbox{$\sim$}}}\ht2=\grdimen\dp2=0pt

\def\dot#1{\vbox{\baselineskip=-1pt\lineskip=1pt
     \halign{\hfil ##\hfil\cr.\cr $#1$\cr}}}

\def\a218{\alpha_{2-18}}

\def\aint218{\alpha_{{\rm int},2-18}}

\def\hexnumber#1{\ifcase#1 0\or1\or2\or3\or4\or5\or6\or7\or8\or9\or
 A\or B\or C\or D\or E\or F\fi }

\makeatletter
\ifx\CUP@mtlplain@loaded\undefined
\else
\fi
\makeatother
 \makeatletter
 \ifx\CUP@mtlplain@loaded\undefined
   \font\tenbmi=cmmib10 at 10pt
   \font\sevenbmi=cmmib10 at 7pt
   \font\fivebmi=cmmib10 at 5pt

   \newfam\bmifam
   \textfont\bmifam=\tenbmi
   \scriptfont\bmifam=\sevenbmi
   \scriptscriptfont\bmifam=\fivebmi
   
 \fi
 \makeatother
\ifnfsstwo

\fi
\ifnfssone

\fi
\ifoldfss

\fi

\mathchardef\varLambda="0103

\makeatletter
\ifx\CUP@mtlplain@loaded\undefined
\else
\fi
\makeatother
\makeatletter
\ifx\CUP@mtlplain@loaded\undefined
  \font\tenbms=cmbsy10
  \font\sevenbms=cmbsy10 at 7pt
  \font\fivebms=cmbsy10 at 5pt
  \newfam\bmsfam
  \textfont\bmsfam=\tenbms
  \scriptfont\bmsfam=\sevenbms
  \scriptscriptfont\bmsfam=\fivebms

  \edef\bsy@{\hexnumber\bmsfam}
  \mathchardef\bnabla="0\bsy@72
\fi
\makeatother
\def\etal{\mbox{\it et al.}}

\title[Advances in accretion]{Old and new advances in black hole 
accretion disc theory}

\author[R. Svensson]%
{R\ls O\ls L\ls A\ls N\ls D\ns  S\ls V\ls E\ls N\ls S\ls S\ls O\ls N}

\affiliation{Stockholm Observatory, S-133 36 Saltsj\"obaden, Stockholm, 
Sweden}

\begin{document}
\ifnfssone
\else
  \ifnfsstwo
  \else
    \ifoldfss
      \let\mathcal\cal
      \let\mathrm\rm
      \let\mathsf\sf
    \fi
  \fi
\fi

\maketitle

\begin{abstract}
A summary is given of the high-lights during the Reykjavik Midsummer
Symposium on Non-Linear Phenomena in Accretion Discs around
Black Holes. Such high-lights include the recent advances on
understanding: 1) the accretion disc solution branch dominated by advection
(i.e., advection dominated accretion flows, ADAFs),
2) the importance of magnetic fields in many different respects, most importantly
being responsible for the self-sustained MHD-turbulence giving rise to the
disc viscosity, and 3) the details of the radiation processes giving
rise to the X/$\gamma$-ray continuum originating close to the black hole.

Some old advances are unfortunately also necessary to discuss here.
It is pointed 
out to the accretion disc research community, that many of the research papers
published on ADAFs 1994-1997 do not accurately present the history of ADAF research.
Some of the results that were presented as new and original actually appeared
in a paper by Ichimaru already in 1977. Also not quoted are the papers
from the 1970's and 1980's calculating the temperature structure of and
the spectra from quasi-spherical accretion
flows onto black holes. As the ADAFs are close to being quasi-spherical,
the resulting spectra of ADAFs and quasi-spherical flows are almost identical.
Further of the recent ADAF results are therefore not new results
as is sometimes claimed.

In spite of all the recent progress of various aspects of accretion flows
around black holes, many of the research lines have still not been 
merged providing potential for further dramatic progress in coming years.
\end{abstract}

\firstsection 

\section{Why Iceland?}
\label{sec:whyiceland}

This is the very first conference in astrophysics that has
ever taken place on Iceland, a country with only about 2-3 astronomers. 
It is natural to ask the question: Why was it organized on Iceland?
This question fortunately has several reasonable answers:

First, the astronomy population on Iceland is not that small as it 
may seem at first sight. It is approximately similar to that in other
Western countries, i.e., about 1 astronomer per 100 000 in population.
Iceland's 250 000 citizens imply a total of 
2.5$ \pm \sqrt{2.5}$ astronomers, in good agreement with 
the actual number.

Second, one of these 
2.5 astronomers works on accretion discs
which makes Iceland probably the country with the world's largest fraction
of astronomers working in this field.

Third, in the past, at least according to classic literature, there has been at least
two historic studies of ``black hole interiors'' originating on Iceland.
In Jules Verne's {\it Voyage au Centre de la Terre} (1864),
the research team led by Professor Otto Lidenbrock from Hamburg 
entered a hole (most likely being black) on the bottom of the
crater of Snaefell in 1863. Here, they followed in the footsteps
of the 16th century Icelandic scientist Arne Saknussemm who had explored this
black hole before them as described in Jules Verne's novel.
 
Fourth, and maybe the most important reason is that the research area of
accretion discs has grown rapidly, not only worldwide, but also
in most of the Nordic countries 
(Denmark, Finland, Iceland, Norway, Sweden) over the last
10 years. Because of this, all the support for this conference
originated from these countries directly or indirectly with the
requirement, of course, that the conference had to take place in one of these
countries.

The summary talk summarized most of the talks at the Midsummer Symposium.
However, as the chapters in this monograph were written up to a year
after the symposium, their content are often different from that 
of the symposium talks occasionally being influenced by the summary talk.  
This summary chapter still summarizes the symposia talks,
but also includes some comments on the material in the written reviews.
 
\section{Accretion discs}
\label{sec:accretiondiscs}

One of the main topics of this Midsummer Symposium is the recent research bandwagon
on advection dominated accretion flows (ADAFs) starting in 1994 with probably 
of the order of 100 papers since then. Many of the results are
covered in Abramowicz (this volume), Bj\"ornsson (this  volume), 
Narayan, Mahadevan, \& Quataert (this volume), and Lasota (this volume). 
The problem is that in all the papers written before
the summary talk of this symposium, 
the history of research on ADAFs has not been accurately presented.
In my summary talk, I wanted to take the opportunity to set
the record straight. I therefore here quote a few paragraphs from
the section on the 
history of theoretical accretion disc research in my chapter (Svensson 1997) in 
{\it Relativistic Astrophysics: A Conference in Honour of Prof.
I.D. Novikov's 60th birthday}:

\subsection{Theoretical history (from Svensson 1997)}
\label{sec:theoryhistory}

The underlying framework for almost all efforts to understand
active galactic nuclei (AGN) is the accretion disc picture 
described in the classical papers by Novikov \& Thorne (1973) and \cite{ss73}.
This original picture was partly inspired by the extreme
optical AGN luminosities.
Here, effectively optically thick, rather cold matter forms a 
geometrically thin,
differentially rotating Keplerian disc around a supermassive 
black hole. The differential motion causes viscous
dissipation of gravitational binding energy
resulting in outward transportation of angular momentum and 
inward transportation of matter.
The dissipated energy diffuses vertically and emerges as black
body radiation, mostly in the optical-UV spectral range for the case of AGN. 

Observations have also been the driving force in
the discovery of two other solution branches.

a) Hard X-rays from the galactic black hole candidate, Cyg X-1,
as well as the discovery of strong X-ray emission from most AGN
led to the need for a hot accretion disc solution.
\cite{sle76} (SLE) found  
a hot, effectively optically thin, rather geometrically thin solution 
branch, where the ions and 
the electrons are in energy balance, with the ions being heated 
by dissipation and cooled
through Coulomb exchange, leading to ion temperatures of order 
$10^{11}$ - $10^{12}$ K. 
The efficient cooling of electrons through a variety of mechanisms
above $10^9$ K, leads to electron temperatures being locked
around $10^9$ - $10^{10}$ K. The SLE-solution is thermally unstable.

b) Observations of Cyg X-1 and of radio galaxies
led to two independent discoveries of the third solution branch.
\cite{ich77} developed a model to explain the soft high and 
hard low state of Cyg X-1. The soft high state is due to the disc being
in the optically thick cold state, and the hard low state
occurs when the disc develops into a very hot, optically thin state. 
This solution branch is similar to  
the SLE-solution, except that now the ions are not in local energy
balance. It was found that if the ions were sufficiently
hot they would not cool on an inflow time scale, but rather
the ions would heat up both by adiabatic compression and viscous dissipation
and would carry most of that energy with them into the black hole.
Only a small fraction would be transferred to the electrons,
so the efficiency of the accretion is much less than the
normal $\sim 10 \%$.
As the ion temperature was found to be close to virial, these discs are 
geometrically  thick. Just as for the SLE-case above, 
the electrons decouple to be locked at  $10^9$ - $10^{10}$ K.
The flows resemble the quasi-spherical dissipative flows studied
by, e.g., \cite{mes75} and \cite{mar82}. 
Independently, \cite{rees82} and \cite{phi83} proposed that a similar
inefficient accretion disc solution is responsible for the low 
nuclear luminosities in radio galaxies with large radio lobes and thus 
quite massive black holes. These 
geometrically  thick discs were named {\it ion tori}.  
\cite{rees82} specified more clearly than \cite{ich77}
the critical accretion rate above which
the flow is dense enough for the ions to cool on an inflow time scale
and the ion tori-branch would not exist. Further
considerations of ion tori were made by \cite{beg87}. 
\cite{ich77} emphasized that his version of ion tori is thermally stable.

The ion-tori branch has been extensively studied 
and applied over the last two years (1995-1996) with more than 30 papers
by Narayan and co-workers, Abramowicz and co-workers, as well as many others. 
These studies confirm and extend the original results, 
although some papers do not quote or recognize the original results.
New terminology has been introduced based on an Eularian viewpoint
rather than a Lagrangian. Instead of the ions not cooling,
it is said that the local volume is "advectively cooled" due to the ions
carrying away their energy. The {\it ion tori} are therefore
renamed as {\it advection-dominated discs}. 

\subsection{Why did history go wrong?}
\label{sec:whywrong}

From the above, it is clear that just 4 years after the start of
research on 
modern accretion disc theory in 1973, the pioneering work on three of the four major
accretion disc branches had been done (see Narayan \etal, this volume for 
a description of the four branches).  
As a graduate student with a side 
interest in accretion discs, I myself read the paper
by Ichimaru back when it was published and a few times since then.
When the ADAF research bandwagon got going in 1995, I noted that 
Ichimaru's pioneering work was not quoted
and that the importance of the work by \cite{rees82} and \cite{phi83} initially
was downplayed. Instead, the ADAF-solution was presented as a
new discovery. Even the issue of whether the ADAF-solution was new or not was discussed
in some papers. Let me give a few quotations: 
``Recently, a new class of two-temperature
advection-dominated solutions has been discovered''; ``We have found new types of
optically thin disc solutions where cooling is dominated by the radial advection of
heat''; ``Is our new solution really new? Certainly, to the extent that we have for the
first time included advection and treated the dynamics of the flow consistently,
the advection-dominated solution... is new. But even more fundamentally, it is our
impression that the existence of {\it two} hot solutions was not appreciated until
now''. 
In order to correct the history of ADAF research, I
wrote the paragraphs above in June 1996 as part of a review on X-rays and gamma-rays
from AGNs (Svensson 1997), submitted it to the pre-print archives, sent preprints to
about 150 astronomical libraries, and spread 100s of copies at conferences. But even
so, the information did not penetrate the research community. So, as a summarizer of
this accretion disc symposium, I took
the liberty to show that the most important results of recent ADAF-research were
already obtained by Ichimaru in 1977. Although Ichimaru's work was not mentioned in
any of the talks at this symposium,  it is gratifying to notice that he is quoted in
three of the chapters. Furthermore, Ichimaru is now regularly quoted in ADAF-papers
appearing since this symposium.

Why was Ichimaru's 1977 paper forgotten by almost everybody?
It may depend on the way literature searches are done. One searches
a few years back, and assumes that the major reviews cover the most
important old papers. And once the ADAF history was written, it became adopted by
the research community. But it does not explain why the Ichimaru-paper
was ignored already in the late 1970s. It was quoted in the Cygnus X-1 review
by Liang \& Nolan (1984) but not for providing a new accretion disc solution,
or for its explanation of the two spectral states of Cyg X-1.

\subsection{ADAF principles}
\label{sec:adafprinciples}

Much of the recent work on different accretion disc solutions
and on ADAFs, in particular, is summarized in Abramowicz (this volume), 
Bj\"ornsson (this volume), and 
Narayan, Mahavedan, \& Quataert (this volume). 
The recent work has among other things developed self-consistent solutions
for the dynamics of ADAFs, has elucidated the connections between the different
solutions branches, and has calculated detailed spectra from such flows.
 
One of the outstanding questions is which of the solution branches (see
Figure 1 in Abramowicz or Bj\"ornsson, this volume) 
a disc chooses. Narayan \& Yi (1995) discussed three possibilities.
In the summary talk, I provided names for these three options 
(and thank Andy Fabian for proposing the ``strong'' and the ``weak'' names).

\begin{itemize}
\item
The Strong ADAF Principle: The disc always chooses the ADAF branch if it exists.
\item
The Weak ADAF Principle: The disc chooses an ADAF branch when no other
stable branch is available.
\item
The Initial Condition Principle: The initial conditions determine which branch
is chosen.
\end{itemize}

One should note here that already Ichimaru (1977) employed the weak and the initial
condition principles in his scenario for the two states of Cyg X-1. 
Depending on the initial conditions at large radii, the disc gas either goes to the
ADAF branch (hard state) or cools down to the gas-pressure dominated 
standard Shakura-Sunyaev branch (soft state). At some smaller radius
in the latter case, radiation pressure
starts dominating, the standard branch becomes unstable, and the weak principle
says that the disc then makes a transition to the ADAF-branch.
Furthermore, Ichimaru (1977) also obtained the result that there is a 
maximum accretion rate above which the optically thin ADAF branch does not
exist, a fact not recognized in recent ADAF-papers where this 
critical accretion rate was rediscovered
(for this accretion rate the cooling time scale becomes equal
to the inflow time scale and the disc cannot remain in an ADAF state,
see \S 3.2.2 in Narayan \etal, this volume).

How unique are the ADAF-models and scenarios that are proposed for various
phenomena?
Depending on which ADAF-principle one uses, one gets different models and scenarios,
which gives some latitude for the model-builder.
It is therefore
important to develop a physical understanding of the transitions
(with changing radius or accretion rate) between the branches 
and to determine which principle applies.
Furthermore, the whole solution structure with its different branches 
(see
Figure 1 in Abramowicz, this volume, or in Bj\"ornsson, this volume)
depends strongly on the type of viscosity prescription that is used.
The prescription mostly used is $t_{r\phi} = \alpha P_{\rm tot}$.
Other prescriptions such as $t_{r\phi} = \alpha P_{\rm gas}$
or  $t_{r\phi} = \alpha \sqrt{P_{\rm rad}P_{\rm gas}}$
are likely to give very different solution structure and scenarios.
This is rarely discussed in the ADAF-literature. Note that the solution 
structure in the Figure quoted only includes bremsstrahlung as a soft photon 
source for Comptonization. The more realistic case of unsaturated
thermal Comptonization (without specifying the soft photon source)
was considered by Zdziarski (1998).

\subsection{Similarities between quasi-spherical accretions and ADAFs}
\label{sec:quasi}

One of the ADAF-results is that the accretion is almost spherical.
As the proton temperature is close to virial, the vertical scale height is
close to the radius, $R$. All physical parameters then scale with radius approximately 
as for free fall spherical accretion (see eqs. 3.15 in  
Narayan \etal, this volume). The coefficients may be different
depending upon the viscosity parameter, $\alpha$, but for $\alpha$ of order unity,
even the coefficients are approximately the same.
This means that the scenarios of dissipative quasi-spherical accretion of the 
late 1970s
(e.g., \cite{mes75} and \cite{mar82}) and ADAFs are essentially identical,
the difference being that the workers of the late 1970s did not consider
the dynamics in detail but rather used intuition to conclude that the
flow scaled as free fall flows. Many of the resulting conclusions are
the same. It was, e.g., noted in some early (quasi-)spherical accretion papers
that the solutions depend on the black hole mass mainly through the combination 
$\dot m \equiv \dot M / \dot M_{\rm Edd} \propto \dot M / M$, and that 
therefore the solutions are
similar for both galactic black holes and for super massive AGN black 
holes. Other results were that a {\it two-temperature} structure develops 
in the inner, say, 100 Schwarzshild radii; that the proton temperature remains close
to virial, while the electron temperature saturates at about a few times
10$^9$ K; and that the luminosity scales as $\dot M^2$. 
Some of the radiation processes that were included were: 
self-absorbed cyclo-synchrotron emission,
Compton scattering, and sometimes pion production in proton-proton collisions. 
The accretion rates, $\dot m$, considered were of order unity
as the purpose was to explain the luminous AGNs.

These results again appear in the ADAF literature (see, e.g.,
Figures 4, 6, and 7 in Narayan \etal, this volume). One important difference
now is that also small $\dot m$  are considered giving rise to very different
spectra (see Figure 6 in the chapter of Narayan \etal, this volume).

\subsection{Applications of ADAFs}
\label{sec:applications}

The ADAFs have  been included in scenarios and models for several astrophysical 
phenomena as described by Narayan \etal\, (this volume) and by Lasota (this volume).
Such phenomena include explaining the quiescent state of three black hole 
X-ray transients, the different spectral states and spectral transitions
of soft X-ray transients,
low-luminosity galactic nuclei such as Sgr A$^*$ in the the Galactic Center,
the LINER NGC 4258, and possible ``dead quasars'' such as  M87 and M60.
Lasota (this volume) discusses in greater detail how the spectral properties 
of the outbursts of the soft X-ray transient GRO J1655-40 can 
be explained within a scenario with an inner ADAF + an outer cold disc. 

As mentioned above, applying different ADAF principles give rise to different 
scenarios. One such case is the efforts to explain the different spectral states
of soft X-ray transients. Chen \& Taam (1996) apply the weak principle and finds
the transition radius between the outer cold disc and the ADAF  to increase with
accretion rate. Esin, McClintock, \& Narayan (1997), on the other hand,
in their detailed work apply the strong ADAF principle and finds the transition
radius to decrease with increasing accretion rate (see Figure 11 in
Narayan \etal, this volume) 

Another case is the LINER NGC 4258 (Narayan \etal, this volume), 
where the strong ADAF principle gives rise to an ADAF inside
the masering molecular disc, while the weak ADAF principle gives rise to a standard
thin disc (Neufeld \& Maloney 1995). Narayan \etal\, (this volume) 
argue that the latter scenario has
a too low accretion rate to explain the observed emission.
 
\section{Evidence for the existence of black holes and surrounding accretion discs}
\label{sec:evidence}

These topics were covered by Andy Fabian in his talk. In this volume,
most of the evidence is, however,  discussed in the two observational chapters
by Charles (Galactic black holes) and by Madejski (supermassive black holes),
while the review chapter by Fabian is limited to the broad Fe emission lines
generated by the inner parts of a cold thin accretion disc.

The evidence for supermassive black holes in galactic nuclei has in the 
past been indirect. X-ray variability has set an upper limit to the size
of the emitting region and thus to the mass.  The luminosity has provided 
an estimate of the mass assuming the object to be radiating at the Eddington
luminosity. Measuring the velocity fields close to the nuclei of nearby galaxies
has also provided an estimate of the mass enclosed within that region.
Recently, there has, however, been dramatic progress as described by 
Madejski (this volume).
The VLBA mega-masers in the LINER NGC 4258 showed  a Keplerian velocity profile
indicating a ``point'' mass of $3.6 \times 10^7 {\rm M}_{\odot}$.
And {\it ASCA} observations showed Fe-line profiles broadened and
distorted in precisely the way expected for emission from a rotating disc 
just outside a black hole (see Figure 4 in Madejski, this volume, or
Figures 5 and 6 in Fabian, this volume). Some observations even sets constraints
on the rotation of the black hole. Future space missions with observing 
capability of the Fe-line will provide ample opportunities to explore the strong
gravitational field close to the event horizon of black holes.
The Fe-line at the same time provides evidence for the existence and the
properties of a cold reflecting disc close to the black hole.

There has been similar progress regarding determining the dynamical mass 
of the compact object in several galactic soft X-ray transients
(see Table 3 in Charles, this volume). Again, the progress is mainly 
observational depending on the X-ray satellites providing the discovery
of the transients, and the very large ground-based telescopes providing
the dynamical mass-determinations. Present and future X-ray missions with
all-sky monitors will discover new transients increasing the statistics
and possibly broadening the range of properties of galactic black holes.

In this context, one should note the exotic but qualitative contribution 
by Novikov (this volume)
on the physics just outside and inside the event horizon. Of particular
interest is the qualitative discussion of the tremendous growth of the internal
mass (mass inflation) of the black hole interior during the formation process.

\section{Radiation processes}
\label{sec:processes}

While the classical rates for bremsstrahlung, cyclo/synchrotron radiation,
and Compton scattering in the nonrelativistic and relativistic limits
were sufficient in the early disc models, the electron temperatures
of $10^9$ - $10^{10}$ K indicated by both observations and theory
required the calculation of transrelativistic rates of the above processes
as well as for pair processes that becomes important at these
temperatures. Rate calculations as well as exploring the properties
of pair and energy balance in hot plasma clouds
 were done in the 1980s by Lightman, Svensson, Zdziarski and
others.  The Compton scattering kernel probably received its definite
treatment in \cite{nag94}. Recent improvements in some rates 
were obtained by Mahadevan, Quataert, and others. Much of this microphysics
have been included both into the SLE-solutions and into the 
ADAF-models. One problem has been to determine the importance of
electron-positron pairs in hot accretion flows. The most 
detailed considerations so far show that pairs have at most
a moderate influence in a very limited region of parameter space
(see Bj\"ornsson, this volume).  
 
To obtain approximate spectra, the Kompaneets equation with 
relativistic corrections and with a simple escape probability replacing the
radiative transfer is sufficient (Lightman \& Zdziarski 1987).
However, if one want to obtain constraints on the geometry from
detailed observed spectra, then methods to obtain exact radiative 
transfer/Comptonization solutions in different accretion geometries 
must be developed. Such methods were developed by
\cite{haa93} (approximate treatment of the Compton scattering)
and \cite{pout96} (exact treatment) who solved the radiative transfer
for each scattering order separately (the iterative scattering method).
These codes are fast enough to be implemented in XSPEC, the standard X-ray
spectral fitting package, and exactly computed model spectra can now be used
when interpreting the observations. The codes, furthermore, includes
the reprocessing (both absorption, reflection, and transmission) 
of X-rays by the cold matter.
These methods have mostly been used to study radiative transfer
in two-phase media consisting of cold and hot gas with simple geometries,
but they have also recently been integrated into the 
ADAF models by Narayan and co-workers.

Poutanen (this volume) concludes that the galactic sources with their smaller
reflection
are best fit with a hot inner disc surrounded by a cold outer disc,
while the Seyfert galaxies with their larger reflection
are best fit with a geometry where 
the X-ray emission originates from active regions (magnetic flares?) atop
of a cold disc (extending all the way in to the black hole as the 
broad, distorted Fe-lines indicates).
The spectral predictions of ADAF models
have not been tested against the detailed spectral observations
using $\chi^2$ fittings. One should therefore at the present moment
have greater
confidence in the conclusions of the work doing detailed radiative transfer and
spectral modeling and fittings in simple geometries. 

Poutanen also describes how the {\it detailed} spectra of the spectral 
transition between  the hard and the soft states of Cyg X-1 can be described
by a simple hybrid pair model (see Figure 9 of Poutanen, this volume) in a
geometry with a hot inner disc surrounded by an outer cold disc and where the
transition radius changes during the transition.
The geometry is similar to the ADAF scenario suggested by Esin \etal\ (see
Figure 14 in Narayan \etal\, this volume).
 
It is clear that the natural evolution is to merge the detailed spectral models
with various accretion disc scenarios. One weakness of the ADAF-literature
is that the spectral predictions of the (broad band) ADAF model is normally only
compared with the standard (narrow band) disc model of Shakura \& Sunyayev (1973). 
Other alternative scenarios,  such as the two phase
scenario where cold clouds are submerged in a  hot medium (e.g.,
Krolik 1998 for a recent work on this scenario), are normally not discussed. 

Another problem is the determination of the
detailed spectral predictions from the cold disc,
where the expected spectrum certainly is not that of a black body 
(Krolik, this volume).

\section{The effects of magnetic fields}
\label{sec:magfields}

Magnetic fields are important in several respects in accretion
discs around black holes. Several of these were
discussed during the symposium and here some of them are listed:
 
\begin{itemize}
\item 
The most important influence of magnetic fields in an accretion disc
is probably as being the agent generating self-sustained 
MHD-turbulence in a differentially rotating disc and thereby 
providing the necessary anomalous viscosity needed in black hole 
accretion discs. Brandenburg (this volume) describes 
the results of 3-D numerical simulations schematically in Figure 2.
The Keplerian shear gives rise to large scale magnetic fields that in its turn 
generates turbulence through the Balbus-Hawley and Parker instabilities.
Finally, the turbulence regenerates the magnetic field through the dynamo 
effect. The energy flow is shown in Figure 5 (Brandenburg, this volume).
Approximately the same power is released in Joule heating (of electrons and ions) 
as in viscous heating (of ions mainly).  One of the most important results 
is the magnitude
of the Shakura-Sunyayev viscosity parameter, $\alpha$, and the realization that
$\alpha$ is not a constant.

\item The magnetic field generates vertical stratification. There will be
two different scale hights for the magnetic and the gaseous pressures.
The magnetic field is more uniformly distributed vertically than the matter.

\item
The magnetic field plays a crucial role in the generation of a
corona around the disc. The hope is to use a short time sequence from
a simulation run of MHD-turbulence in the disc itself.
This time sequence will form the driving background for a corona simulation.
Nordlund speculated that spontaneous  magnetic dissipation
may generate self-organized criticality in the corona, similar
to what has been found in simulations for the solar corona.
Here, one should also note the partly unsuccessful efforts described
by Wiita (this volume) in obtaining a self-organized critical state
of the accretion disc itself.

\item
The magnetic field may act as a confining agent for cold gas (in the scenario 
where cold gas clouds coexist with a hot teneous medium) as discussed 
by Celotti \& Rees (this volume). The field may also be responsible for confining
the matter or pairs in the active regions discussed by Poutanen (this volume).

\item
The magnetic field plays an important role as electrons (and positrons)
in its presence generates cyclo-synchrotron photons in the radio to 
IR-range. These photons serve as seed photons in the Comptonization process,
and is a crucial component in the ADAF models (Narayan \etal, this volume).

\item
Under certain conditions, synchrotron self-absorption dominates
over Coulomb scattering as a thermalizing mechanism.
The electrons emit and absorb cyclo-synchrotron photons
resulting in
a Rayleigh-Jeans self-absorbed photon spectrum and a 
Maxwellian distribution for the electrons
(Ghisellini, Guilbert, \& Svensson 1988;
Ghisellini \& Svensson 1989; Ghisellini, Haardt, \& Svensson 1998).
The thermalization time scale is just a few synchrotron cooling times.

 \end{itemize}

\section{Larger scale phenomena}
\label{sec:largescale}

Some phenomena on larger scales were discussed during the symposium.
In an interesting talk, Pringle showed how the radiation from the
central source may cause radiation-driven warping of the disc on
larger scales. In certain cases, the inner disc may even turn upside 
down relative the outer disc resulting in self-shadowing of the radiation
from the central source. The resulting ionization cone may not necessarily 
be aligned with the central source. 

Merging of galaxies containing supermassive black holes may
lead to a supermassive binary black hole in the resulting galaxy.
The question is: How does the binary black hole evolve?
Artymowicz (this volume) discusses the history of the efforts trying to
solve this problem. After some 20 years of research it now seems that dynamical
friction does not cause the eccentricity of the black hole orbits to grow.
The interaction of a binary black hole with a common disc may resolve 
the difficulty of getting black hole-black hole merging to occur in
less than a Hubble time. In the process, the black holes are fed, and 
a periodic light curve may be observed. One such example is OJ 287
with a period of about 13 years.
A black hole circling a primary black hole with an accretion disc
may cause warping of the disc as shown by Papaloizou \etal\, (this volume).
The obvious application is the observed warping of the masering disc
in the LINER NGC 4258.

Another issue discussed was the survival of vortices in accretion discs.
The common notion is that Keplerian shear would kill such vortices 
on a few rotation time scales. In Spiegel's talk (and in Bracco \etal,
this volume)  it was shown that coherent structures indeed form.
On the other hand, in the local simulations by Brandenburg (this volume)
including magnetic fields, vortices do not form. The issue of vortices, 
their survival and influence is not yet settled.

\section{Conclusion}
\label{sec:conclusions}

This was a most rewarding symposium where leading scientists
presented the most recent dramatic developments regarding several of
the physics areas needed for realistic modelling
of black hole accretion flows.
These areas include radiative processes in hot plasmas,
radiation transfer in hot and cold plasmas,  
MHD in differentially rotating gas, the origin of disc viscosity,
magnetic flares, gas or MHD-simulations of flows, and so on.
Some of these research lines have already merged.
As each subproblem is understood, further merging provides
the potential for dramatic progress in coming years.

\begin{acknowledgments}
      The author acknowledges support from the
      Swedish Natural Science Research Council and the Swedish National 
      Space Board. 
\end{acknowledgments}

\end{document}